\documentstyle[12pt]{article}
 
\title{Normal ordering for deformed boson operators and
operator-valued deformed Stirling numbers}
\author{ Jacob Katriel\thanks{Permanent address:
Department of Chemistry, Technion - Israel Institute of Technology,
Haifa 32000, Israel. email: chr09kt@technion.}
$\; $ and Maurice Kibler\thanks{email: Kibler@ipnl.in2p3.fr.} \\
{\small \sl Institut de Physique Nucl\'{e}aire de Lyon } \\
{\small \sl IN2P3-CNRS et Universit\'{e} Claude Bernard} \\
{\small \sl 43 Boulevard du 11 Novembre 1918} \\
{\small \sl F-69622 Villeurbanne Cedex, France} }
 
\date{ }
\begin{document}
\setcounter{page}{1}
 
\maketitle
 
\begin{abstract}
 
The normal ordering formulae for powers of the boson number operator
$\hat{n}$
are extended to deformed bosons. It is found that for the ``M-type''
deformed bosons, which satisfy $a a^{\dagger} - q a^{\dagger} a = 1$,
the extension involves a set of deformed Stirling numbers which
replace the Stirling numbers occurring in the conventional case.
On the other hand, the deformed Stirling numbers which
have to be introduced in the case of the ``P-type'' deformed bosons,
which satisfy $a a^{\dagger} - q a^{\dagger} a = q^{-\hat{n}}$,
are found to depend on the operator $\hat{n}$. This distinction
between the two types of deformed bosons is in harmony with earlier
observations made in the context of a study of the extended
Campbell-Baker-Hausdorff formula.

Published in J.~Phys.~A~:~Math.~Gen.~25 (1992) 2683-2691. 
 
\end{abstract}
 
\newpage
 
\section{Introduction}

\indent
The transformation of a second-quantized operator
into a normally ordered form, in which each
term is written with the creation operators preceding the annihilation
operators, has been found to simplify quantum mechanical calculations in
a large and varied range of situations. Techniques for the
accomplishment of this ordering  have been developed and are widely
utilized~[1,2]. A particular
subclass of problems and techniques involves situations in which the
operators of interest commute with the number operator.
More specifically,
one is interested in transforming an operator  which is  a function  of
the number operator into a normally ordered form,
or transforming an operator
each of whose terms has an equal number of creation and annihilation
operators corresponding to each degree of freedom, into an equivalent
operator expressed in terms of the number operator only.
 
In the present article we consider the corresponding problem for the
deformed bosons which have been investigated very extensively in the
last three years~[3,4]
in connection with the recent interest in
the properties and
applications of quantum groups.
 
\section{Stirling and deformed Stirling numbers}
 
\indent
The Stirling numbers of the first $(s)$
and second $(S)$ kinds were introduced in
connection with the expression for a descending product of a
variable $x$ as a linear combination of integral and positive powers of
that variable, and the inverse relation, respectively~\cite{Abra}
\begin{equation}
x \; (x-1) \; \cdots \; (x-k+1) = \sum_{m=1}^k \; s(k,m) \; x^m
\label{eq:e1}
\end{equation}
\begin{equation}
x^m = \sum_{k=1}^m S(m,k) \; x \; (x-1) \; \cdots \; (x-k+1) \; .
\label{eq:e2}
\end{equation}
\noindent
Using these defining relations it is easy to show that the Stirling
numbers satisfy the recurrence relations
\begin{equation}
s(k+1,m)=s(k,m-1)-k \; s(k,m)
\label{eq:e3}
\end{equation}
\noindent
and
\begin{equation}
S(m+1,k)=S(m,k-1)+ k \; S(m,k) \; ,
\label{eq:e4}
\end{equation}
\noindent
with the initial values $s(1,1)=S(1,1)=1$
and the ``boundary conditions''
$ s(i,j)=S(i,j)=0$ for
$i<1$, $j<1$ and for $i<j$.
The combinatorial significance of the Stirling numbers has been amply
discussed~\cite{Riordan}.
 
Several generalisations of the Stirling numbers appeared in the
mathematical literature [7-12].
In anticipation of further development we shall refer to them
generically as deformed Stirling numbers. In this context
we wish to distinguish between
the two widely used forms of ``deformed numbers''
$[x]_M=\frac {q^x -1}{q-1}$, the usual choice in the
mathematical literature on $q$-analysis \cite{Exton}, and
$[x]_P = \frac{q^x -q^{-x}}{q-q^{-1}} $,
which is common to the recent physical literature and to
the literature on quantum groups.
A generalisation was recently proposed by
Wachs and White \cite {Wachs}, which can be written in the form
$[x]_G =\frac {q^x-p^x}{q-p}$.
This form contains $[x]_M$ and $[x]_P$ as special cases,
corresponding to the choices $p=1$ and $p=q^{-1}$, respectively.
We shall write $[x]_{M(q)}$, $[x]_{P(q)}$ and
$[x]_{G(p,q)}$ instead of the symbols introduced above
whenever the choice of the parameters $q$ and/or $p$ will
have to be explicated.
The identities
\begin{equation}
[x]_{P(q)}= q^{-x+1} \; [x]_{M(q^2)} \qquad
[x]_{G(p,q)}= p^{x-1} \; [x]_{M(q/p)} \qquad
[x]_{G(p,q)}=(\sqrt {pq})^{x-1} \; [x]_{P(\sqrt{q/p})}
\label{eq:EQ5}
\end{equation}
\noindent
illustrate the notation and exhibit
some of the elementary properties of these deformed numbers.
 
One of the generalisations of the Stirling numbers \cite{Milne} involves
a descending product of M-type deformed
numbers expressed in terms of the powers
of the M-type deformed number $[x]_M$
\begin{equation}
[x]_M \; [x-1]_M \; \cdots \; [x-k+1]_M =
\sum_{m=1}^k \; s_q(k,m) \; [x]_M^m
\label{eq:e5}
\end{equation}
\noindent
and the corresponding inverse relation
\begin{equation}
[x]_M^m = \sum_{k=1}^m \;
S_q(m,k) \; [x]_M \; [x-1]_M \; \cdots \; [x-k+1]_M \; .
\label{eq:e6}
\end{equation}
\noindent
Using the defining relations it is easy
to show that the deformed Stirling numbers $s_q(k,m)$ and $S_q(m,k)$,
which are referred to in the mathematical literature as $q$-Stirling
numbers of the first and second kind, respectively,
satisfy the recurrence relations
\begin{equation}
s_q(k+1,m)=q^{-k} \; \Big{(} s_q(k,m-1)-[k]_M \; s_q(k,m) \Big{)}
\label{eq:e7}
\end{equation}
\noindent
and
\begin{equation}
S_q(m+1,k)=q^{k-1} \; S_q(m,k-1)+ [k]_M \; S_q(m,k) \; ,
\label{eq:e8}
\end{equation}
\noindent
with ``boundary conditions'' and initial values identical with those
specified above for the conventional Stirling numbers.
 
A slight modification in the form of the descending product, replacing
the factors $[x-i]_M$ by $[x]_M - [i]_M$, results in the relations [8-10]
\begin{equation}
[x]_M \; ([x]_M-[1]_M) \; \cdots \; ([x]_M-[k-1]_M) =
 \sum_{m=1}^k \; \tilde{s}_q(k,m) \; [x]_M^m
\label{eq:e9}
\end{equation}
\noindent
and
\begin{equation}
[x]_M^m = \sum_{k=1}^m \; \tilde{S}_q(m,k)
\; [x]_M \; ([x]_M-[1]_M) \; \cdots \; ([x]_M-[k-1]_M) \; .
\label{eq:e10}
\end{equation}
\noindent
Starting with these defining relations and using the identity
\cite{Gould}
\begin{equation}
[a]_M-[b]_M=q^b[a-b]_M
\label{eq:iden11}
\end{equation}
we obtain the recurrence relations
\begin{equation}
\tilde {s}_q(k+1,m)= \tilde{s}_q(k,m-1)-[k]_M \; \tilde{s}_q(k,m)
\label{eq:e11}
\end{equation}
\noindent
and
\begin{equation}
\tilde{S}_q(m+1,k)=
\tilde{S}_q(m,k-1)+ [k]_M \; \tilde{S}_q(m,k) \; ,
\label{eq:e12}
\end{equation}
\noindent
where the ``boundary conditions'' and initial values are, again,
as above. Note that
\begin{equation}
\tilde{s}_q(k,m)=q^{k(k-1)/2} \; s_q(k,m) \qquad \qquad \qquad
\tilde{S}_q(m,k)=q^{-k(k-1)/2} \; S_q(m,k) \; .
\label{eq:E12}
\end{equation}
\indent
The two sets of deformed Stirling numbers of the first and second
kinds,
as well as the conventional Stirling numbers to which they
reduce in the limit $q \rightarrow 1$, satisfy the following
dual relations
\begin{equation}
   \sum_{m=1}^k s_q(k,m) \; S_q(m,k^{\prime}) =
   \delta(k,k^{\prime})
\label{eq:F12}
\end{equation}
\noindent
and
\begin{equation}
 \; \;  \sum_{k=1}^m S_q(m,k) \; s_q(k,m^{\prime}) =
   \delta(m, m^{\prime}) \; .
\label{eq:G12}
\end{equation}
\indent
An additional set of deformed Stirling numbers of the second kind
was recently introduced by Wachs and White~\cite{Wachs}.
Their definition is motivated by combinatorial
considerations and has no algebraic origin. Their recurrence
relation reads
\begin{equation}
S_{p,q}(m+1,k)= p^{k-1} \; S_{p,q}(m,k-1) + [k]_G \; S_{p,q}(m,k)
\label{eq:Wachs}
\end{equation}
\noindent
and it reduces to (\ref{eq:e12}) for $p=1$.
 
\section{Some algebraic properties of deformed boson operators}
 
\indent
In the context of recent interest in quantum groups and their
realization, three types of
deformed boson operators have
been introduced [3,4,14]. The most straightforward definition starts by
postulating a Fock space on which creation ($a$), annihilation
($a^{\dagger}$)
and number ($\hat{n}$) operators are
defined in analogy with the conventional boson operators. The general
form postulated is
\begin{equation}
a \; |l> = \sqrt {[l]} \; |l-1> \qquad
a^{\dagger} \; |l> = \sqrt {[l+1]} \; |l+1>
\qquad \hat{n} \; |l>=l \; |l> \; .
\label{eq:H12}
\end{equation}
\noindent
It follows immediately that $a^{\dagger}a=[\hat{n}]$ and
$a a^{\dagger}=[\hat{n}+1]$.
The two widely used forms of the deformed
bosons are obtained by choosing either
$[l]=[l]_M=\frac {q^l -1}{q-1}$ or
$[l]=[l]_P = \frac{q^l -q^{-l}}{q-q^{-1}} $.
A generalisation was recently proposed by
Chakrabarti and Jagannathan~\cite{Chak}. We shall adhere to
the notation introduced by Wachs and White~\cite {Wachs} and
write this generalisation in the form
$[l]=[l]_G =\frac {q^l-p^l}{q-p}$, which is
trivially modified relative to that introduced in
Ref.~\cite{Chak}.
As a consequence of a remark made in the previous section, this
third type of deformed boson contains the first two as special cases.
 
The deformed bosons as defined by Eq.~(\ref{eq:H12}) are not associated
with any a priori specification of a (possibly deformed)
commutation relation. Choosing a parameter $Q$, which does not have to
be related to the two parameters $p$ and $q$ so far introduced,
the deformed bosons are found to satisfy the deformed commutation
relation
\begin{equation}
[a, a^{\dagger} ]_Q = a a^{\dagger} - Q a^{\dagger} a =\phi (\hat{n})
=\frac{1}{q-p} \Big{(} q^{\hat{n}}(q-Q) + p^{\hat{n}}(Q-p) \Big{)} \; .
\label{eq:e13}
\end{equation}
\noindent
Since the choice of $Q$ is arbitrary we can opt to be guided by the
requirement that the form of $\phi(\hat{n})$ be as simple as possible
or by some other relevant criterion.
The conventional choice $Q=q$, to which we will eventually adhere,
results in
\begin{equation}
a \; a^{\dagger} - q\; a^{\dagger} \; a = \phi_M(\hat{n})= 1 \; ,
\label{eq:MT}
\end{equation}
\begin{equation}
a \; a^{\dagger} - q\; a^{\dagger} \; a = \phi_P(\hat{n})= q^{-\hat{n}}
\label{eq:P}
\end{equation}
\noindent
and
\begin{equation}
a \; a^{\dagger} - q \; a^{\dagger} \; a =
\phi_G(\hat{n})= p^{\hat{n}}
\label{eq:GT}
\end{equation}
\noindent
for the M-type, P-type and G-type bosons, respectively. We do not
label the creation and annihilation operators by indices such
as M, P or G because the nature of these operators is always
obvious from the context.
The choice $Q=p$ results in
$\phi(\hat{n})=q^{\hat{n}}$ for all the three cases.
For the M-type bosons ($p=1$) this choice implies $Q=1$, i.e.,
the deformed commutation relation becomes
$a \; a^{\dagger} - a^{\dagger} \; a = q^{\hat{n}}$.
For the P-type bosons ($p=q^{-1}$) this choice is the familiar
alternative to Eq.~(\ref{eq:P}), namely
$a \; a^{\dagger} - q^{-1} a^{\dagger} \; a = q^{\hat{n}}$.
In a recent study of the extension of the
Campbell-Baker-Hausdorff formula to deformed bosons \cite{Katsol}, it was
noted that the choice $Q=q$ is the most suitable one for
the M-type bosons, but that $Q=q+q^{-1}-1$ seems to have some advantages
for the P-type bosons. From the same point of view, one would
choose $Q=q+p-1$ for the G-type bosons.
 
We shall also need the relation
\begin{equation}
[a^k, a^{\dagger}]_{Q^k} = \Phi (k,\hat{n}) \; a^{k-1}
\label{eq:e14}
\end{equation}
\noindent
which can be viewed as an extension of Eq.~(\ref{eq:e13}) in the sense
that $\Phi(1,\hat{n})=\phi(\hat{n})$.
One easily finds that
\begin{equation}
\Phi(k,\hat{n})=\frac{1}{q-p} \Big{(}
q^{\hat{n}} \; (q-Q) \; [k]_{G(Q,q)} +
p^{\hat{n}} \; (Q-p) \; [k]_{G(Q,p)} \Big{)} \; .
\label{eq:GenPhi}
\end{equation}
\noindent
We shall retain the conventional choice $Q=q$
for the three cases specified above.
With this choice we get
\begin{equation}
\Phi_M(k,\hat{n})=[k]_M  \qquad
\Phi_P(k,\hat{n})=[k]_P \; q^{-\hat{n}}  \qquad
\Phi_G(k,\hat{n})=[k]_G \; p^{\hat{n}} \; .
\label{eq:Phis}
\end{equation}
 
\section{Normal ordering of powers of the deformed number operator}
 
\indent
The relevance of the ordinary Stirling numbers to the normal ordering of
powers of the boson number operator was demonstrated in
Ref.~\cite{Katriel}.
In the present section we consider some normal ordering properties
of the deformed bosons specified by the parameter choice $p=1$ and
$Q=q$, which corresponds to the M-type boson operators
and to the deformed commutation relation~(\ref{eq:MT}).
Up to a trivial interchange of $p$ and $q$
this is the only combination of parameters for which the
deformed commutator does not depend on $\hat{n}$.
The other types of deformed boson operators are considered in the
following
section where it is found that they differ in a significant respect from
the case presently considered.
 
In order to express an integral power of $[\hat{n}]_M$
in a normally ordered form we can either formally write
such an expansion and obtain a recurrence relation for
the coefficients by applying Eq.~(\ref{eq:MT}) or
use the deformed Stirling
numbers of the second kind directly. We shall present
both approaches because of the intrinsic interest of each one of them.
 
In the direct approach, we start from the expansion
\begin{equation}
[\hat{n}]_M^m=
(a^{\dagger}a )^m = \sum_{k=1}^m \; c(m,k) \; (a^{\dagger} )^k a^k \; .
\label{eq:e16}
\end{equation}
\noindent
Expressing $(a^{\dagger}a)^{m+1}$ by means of Eq.~(\ref{eq:e16}) and
using Eq.~(\ref{eq:MT}), we obtain
a recurrence relation
which is identical with the one satisfied by $S_q(m,k)$,
Eq.~(\ref{eq:e8}). Moreover,
it is obvious from the defining equation~(\ref{eq:e16}) that
$c(1,1)=S_q(1,1)=1$. Thus, $c(m,k)=S_q(m,k)$.
 
A different derivation can be obtained by using the identity
\begin{equation}
\prod_{i=0}^{k-1} \; [\hat{n}-i]_M = (a^{\dagger})^{k} a^{k} \; .
\label{eq:e17}
\end{equation}
\noindent
This identity follows by noting that application of both sides of
Eq.~(\ref{eq:e17}) on any member of the complete set
$\{|l> \, ; \, l=0, \, 1, \, \cdots \}$
of eigenstates of the number operator
results in $ \prod_{i=0}^{k-1} \; [l-i]_M $.
Using Eq.~(\ref{eq:e6}) we obtain
\begin{equation}
[\hat{n}]_M^m = \sum_{k=1}^m S_q(m,k) \prod
_{i=0}^{k-1} \; [\hat{n} - i]_M
\label{eq:hat}
\end{equation}
\noindent
and substituting Eq.~(\ref{eq:e17}) we get the desired normally ordered
expansion
\begin{equation}
[\hat{n}]_M^m = \sum_{k=1}^m \; S_q(m,k) \; (a^{\dagger})^k a^k \; .
\label{eq:E17}
\end{equation}
\indent
We note in passing that an equivalent expansion could have been obtained
starting from the identity
\begin{equation}
\prod_{i=0}^{k-1} \; ( [\hat{n}]_M  - [i]_M ) =
q^{k(k-1)/2} \; (a^{\dagger})^{k} a^{k} \; .
\label{eq:e18}
\end{equation}
\noindent
This identity can be proved either
by induction or by considering the effect
of both sides on the complete set of eigenstates of the number operator.
Using~(\ref{eq:e10}) and (\ref{eq:e18}), we obtain the normally ordered
expansion of $[\hat{n}]_M^m$
in the form
\begin{equation}
[\hat{n}]_M^m = \sum_{k=1}^m \;
\tilde {S}_q (m,k) \; q^{k(k-1)/2} \; (a^{\dagger})^k \;  a^k
\label{eq:e19}
\end{equation}
\noindent
which is related to~(\ref{eq:E17}) by Eq.~(\ref{eq:E12}).
 
In order  to obtain the inverse relation,
expressing a normally ordered product
as a function of the number operator, we note that 
Eqs.~(\ref{eq:e5}) and~(\ref{eq:e17}) lead to
\begin{equation}
(a^{\dagger})^k a^k=[\hat{n}]_M \; [\hat{n}-1]_M \cdots [\hat{n}-k+1]_M
=\sum_{m=1}^k \; s_q(k,m) \; [\hat{n}]_M^m \; .
\label{eq:e29}
\end{equation}
 
\section{Operator-valued deformed Stirling numbers}
 
\indent
In the present section, we attempt to derive
the normally ordered expansion of a power of the number operator for
arbitrarily deformed bosons.
Allowing $p$, $q$ and $Q$ to be arbitrary,
we demand
\begin{equation}
[\hat{n}]_G^m= \sum_{k=1}^m \;
(a^{\dagger})^k \; \hat{S}(m,k,\hat{n}) \; a^k \; .
\label{eq:e21}
\end{equation}
\noindent
Using the general relation (\ref{eq:e14}),
we derive the recurrence relation
\begin{equation}
\hat{S}(m+1,k,\hat{n}) =
Q^{k-1} \; \hat{S}(m,k-1,\hat{n}+1) + \hat{S}(m,k,\hat{n}) \;
\Phi(k,\hat{n}) \; .
\label{eq:e22}
\end{equation}
\noindent
The ``boundary conditions'' and initial values, for all values of
$\hat{n}$, are the same as those following Eq.~(\ref{eq:e4}).
 
The M-type bosons ($p=1$), with the choice $Q=q$ which yields
$ \Phi_M(k,\hat{n}) = [k]_M$, were studied in section~4.
For this case, $\hat{S}(m,k,\hat{n})$ does not depend on
$\hat{n}$. More specifically,
Eq.~(\ref{eq:e22}) then reduces
to Eq.~(\ref{eq:e8}). For the
G-type bosons, we found in section~3 that by choosing $Q=q$ we
obtain $ \Phi_G(k,\hat{n})=[k]_G \; p^{\hat{n}} $ ; consequently,
we have
\begin{equation}
\hat{S}_G(m+1,k,\hat{n})=
q^{k-1} \; \hat{S}_G (m,k-1,\hat{n}+1) +
\hat{S}_G (m,k,\hat{n}) \; [k]_G \; p^{\hat{n}} \; .
\label{eq:e23}
\end{equation}
\noindent
Note that in the general case
$\hat{S}_G(m,k,\hat{n})$ depends on the operator $\hat{n}$.
The special cases $p=1$ and $p=q^{-1}$ are contained in
Eq.~(\ref{eq:e23}).
The dependence of $\hat{S}_G(m,k,\hat{n})$ on $\hat{n}$ for all cases
except $p=1$ can be taken to imply that we have
actually failed to obtain a normally
ordered expansion for $[\hat{n}]_G^m$ in terms of a finite sum in
$ (a^{\dagger})^k a^k$ with $k=1, \, 2, \, \cdots , \, m $.
 
The structure of the recurrence  relation~(\ref{eq:e23}) indicates
that the dependence on $\hat{n}$ of the deformed Stirling
numbers $\hat{S}_G (m,k,\hat{n})$
can be expressed in terms of the factor
$p^{(m-k)\hat{n}}$. Defining the ($\hat{n}$-independent) reduced
Stirling numbers of the second kind
$\Xi (m,k)$ through
\begin{equation}
\hat{S}_G (m,k,\hat{n})=
q^{k(k-1)/2} \; p^{(m-k)\hat{n}} \; \Xi (m,k)
\label{eq:Sigma}
\end{equation}
\noindent
we obtain the recurrence relation
\begin{equation}
\Xi(m+1,k)=p^{m-k+1} \; \Xi(m,k-1) + [k]_G \; \Xi(m,k)
\label{eq:Xi}
\end{equation}
\noindent
with  the initial condition $\Xi(1,1)=1$.
 
To obtain
the ``inverse relation'' to~(\ref{eq:e21}),
expressing a normally ordered term
$(a^{\dagger})^k a^k$ by means of a polynomial in $[\hat{n}]_G$,
we need the ``G-arithmetic'' identity
\begin{equation}
[a-b]_G=q^{-b} ( [a]_G - p^{a-b} [b]_G ) \; ,
\label{eq:f24}
\end{equation}
\noindent
which follows from the two identities
\begin{equation}
[a+b]_G = q^b [a]_G + p^a [b]_G
\label{eq:g24}
\end{equation}
\noindent
and
\begin{equation}
[-b]_G = - (pq)^{-b} \; [b]_G \; .
\label{eq:h24}
\end{equation}
\noindent
We now proceed to obtain the desired relation
\begin{equation}
(a^{\dagger})^k a^k = \sum_{m=1}^k \hat{s}_G (k,m,\hat{n})
 \; [\hat{n}]_G^m \; .
\label{eq:i24}
\end{equation}
\noindent
Since  $(a^{\dagger})^{k+1} a^{k+1} =
(a^{\dagger})^k \; [\hat{n}]_G \; a^k
= (a^{\dagger})^k \; a^k \; [\hat{n}-k]_G$,
we can use Eqs.~(\ref{eq:f24}) and (\ref{eq:i24}) to obtain
the recurrence relation
\begin{equation}
\hat{s}_G(k+1,m,\hat{n}) = q^{-k} \Big{(} \hat{s}_G(k,m-1,\hat{n}) -
p^{\hat{n}-k} \; [k]_G \;
\hat{s}_G (k,m,\hat{n}) \Big{)} \; .
\label{eq:j24}
\end{equation}
\noindent
Note that for $p=1$ this recurrence relation reduces to
Eq.~(\ref{eq:e7}).
 
Introducing the ($\hat{n}$-independent) reduced Stirling numbers
of the first kind
$\xi(k,m)$ such that
\begin{equation}
\hat{s}_G (k,m,\hat{n})= q^{-k(k-1)/2} \; p^{(k-m)\hat{n}} \;
\xi(k,m)
\end{equation}
\noindent
in Eq.~(\ref{eq:j24}), we obtain the recurrence relation
\begin{equation}
\xi (k+1,m) = \xi (k,m-1) - p^{-k} \; [k]_G \; \xi(k,m) \; .
\end{equation}
 
The exponential dependence on $\hat{n}$ of the
deformed Stirling numbers of the first kind, $\hat{s}_G(k,m,\hat{n})$,
means that we have not been able to
express $(a^{\dagger})^k a^k$ as
a {\it polynomial} in $\hat{n}$ but
we did express it as a
{\it function} of $\hat{n}$.
 
In order to derive the bi-orthogonality relations between the
deformed Stirling numbers of the first and second kinds, we first
rewrite Eq.~(\ref{eq:e21}) in the form
\begin{equation}
[\hat{n}]_G^m= \sum_{k=1}^m \;
(a^{\dagger})^k \; a^k \; \hat{S}_G (m,k,\hat{n}-k) \; .
\label{eq:et21}
\end{equation}
\noindent
Using Eq.~(\ref{eq:Sigma}) we obtain
\begin{equation}
\hat{S}_G(m,k,\hat{n}-k) = p^{k(k-m)} \hat{S}_G(m,k,\hat{n}) \; .
\end{equation}
\noindent
Defining $\Xi^{\prime}(m,k) = p^{k(k-m)} \Xi(m,k)$,
we obtain relations of the form of Eqs.~(\ref{eq:F12}) and~(\ref{eq:G12})
with $\Xi^{\prime}(m,k)$ replacing $S_q(m,k)$
and $\xi(k,m)$ replacing $s_q(k,m)$.
 
\section{A generating function for the deformed Stirling numbers
of the first kind}

We start by transforming the $q$-binomial theorem \cite{Exton} into a
G-binomial theorem. By introducing the symbol
\begin{equation}
(\lambda ;x)^{(l)}= (\lambda +x) \; (p\lambda +qx) \;
(p^2\lambda +q^2x) \cdots
(p^{l-1}\lambda +q^{l-1}x)
\label{eq:bin0}
\end{equation}
\noindent
we have
\begin{equation}
(\lambda ;x)^{(l)}=
\sum_{i=0}^l \left[\matrix{&l& \cr
                           &i& \cr } \right]_G \;
p^{i(i-1)/2}
\; q^{(l-i)(l-i-1)/2} \; \lambda ^i \; x^{l-i} \; ,
\label{eq:bin}
\end{equation}
\noindent
where
\begin{equation}
 \left[\matrix{&l& \cr
               &i& \cr } \right]_G
= \frac{[l]_G!}{[i]_G! \; [l-i]_G!}
\end{equation}
\noindent
is a G-binomial coefficient and
$[k]_G!=[1]_G [2]_G \cdots [k]_G$.
Equation~(\ref{eq:bin}) can be proved by induction, using the
G-binomial coefficient recurrence relation
\begin{equation}
 \left[\matrix{l&+&1 \cr
       \hfill   &i& \hfill  \cr } \right]_G =
p^{\; l+1-i}
 \left[\matrix{ \hfill  &l&  \hfill \cr
                       i&-&1   \cr } \right]_G +
q^{i}
 \left[\matrix{   &l&  \cr
                  &i&   \cr } \right]_G \; ,
\end{equation}
\noindent
which follows from the definition of the G-binomial coefficient
on using the G-arithmetic relation~(\ref{eq:g24}).
 
Now, from the identity
\begin{equation}
\frac{(a^{\dagger})^k a^k}{[k]_G!} \; |l> =
 \left[\matrix{&l& \cr
               &k& \cr } \right]_G   |l>
\end{equation}
\noindent
we obtain
\begin{equation}
\sum_{k=0}^{m} p^{k(k-1)/2} \; q^{(l-k)(l-k-1)/2} \; \lambda ^k
\; \frac{(a^{\dagger})^k a^k}{[k]_G!} \; |l> =
(\lambda ; 1)^{(l)} \; |l>
\end{equation}
\noindent
which can be written as an operator identity
\begin{equation}
\sum_{k=0}^{\infty} p^{k(k-1)/2} \;
q^{(\hat{n}-k)(\hat{n}-k-1)/2} \; \lambda ^k \;
\frac{(a^{\dagger})^k a^k}{[k]_G!}  =
(\lambda ; 1)^{(\hat{n})} \; .
\label{eq:iden}
\end{equation}
\noindent
To obtain an expression for $(a^{\dagger})^k a^k$ as a function of
the number operator $\hat{n}$, we have to expand the right-hand side
of Eq.~(\ref{eq:iden}) in powers of $\lambda$.
The coefficient of $\lambda ^k$ can be extracted by writing
\begin{equation}
(a^{\dagger})^k a^k =
\frac {[k]_G!}{k!} \; p^{-k(k-1)/2} \; q^{-(\hat{n}-k)(\hat{n}-k-1)/2}
\; \frac{\partial ^k}{\partial \lambda ^k} (\lambda; 1)^{(\hat{n})}
\bigg {|}_{\lambda =0} \; .
\label{eq:beauty}
\end{equation}
\noindent
The identities
\begin{equation}
[m]_{G(p^k,q^k)} = \frac{[km]_{G(p,q)}} {[k]_{G(p,q)} }
\label{eq:nice}
\end{equation}
\noindent
and
\begin{equation}
[km]_{G(p,q)} = \sum_{i=1}^k
 \left(\matrix{&k& \cr
               &i& \cr } \right)
\; (q-p)^{i-1} \; [m]_{G(p,q)}^i \; p^{m(k-i)}
\label{eq:nicer}
\end{equation}
\noindent
are found to be useful when implementing Eq.~(\ref{eq:beauty}).
(To avoid possible confusion we point out that the symbol
appearing in Eq.~(\ref{eq:nicer}) is the conventional
binomial coefficient.)
Note that for the conventional bosons, for which $p=q=1$,
Eq.~(\ref{eq:beauty}) reduces to an expression~\cite{Katriel}
which can be related to the well-known generating
function for the conventional Stirling numbers of the first
kind~\cite{Abra}.
 
 
\section{Discussion}
 
\indent
In the present article we found that the normal ordering formulae
for powers of the boson number operator can be extended in a
simple and natural way to the M-type bosons, which satisfy
$[a, a^{\dagger}]_q=1$. However, for the P-type bosons,
which satisfy
$[a, a^{\dagger}]_q=q^{-\hat{n}}$, as well as for the more general
G-type bosons, we found that the extension of the conventional
boson analysis results in ``normal-ordering'' expressions with
$\hat{n}$-dependent coefficients.
 
The marked difference between the M-type bosons and all the others
has already been noted before, in the context of the extension of the
Campbell-Baker-Hausdorff formula for products of exponential
operators~\cite{Katsol}. While the observations pointed out
above set apart the M-type bosons, the following may be taken to set
apart the P-type bosons disfavourably,
within the general set of G-type bosons:
Taking the Hamiltonian of the deformed harmonic oscillator to be
${\cal H}=\frac{\hbar \omega_0}{2} (a^{\dagger}a + a a^{\dagger})$
and expanding in
powers of $s=\ln{q}$ and $t=\ln{p}$ (which we assume to be sufficiently
small), we find that
\begin{equation}
{\cal H}=\hbar \omega_0 \lbrack \frac{s+t}{8}
\; + \; (1-\frac{s+t}{2})(\hat{n}+\frac{1}{2}) \; + \;
\frac{s+t}{2} (\hat{n}+\frac{1}{2})^2+\cdots \; \rbrack \; .
\label{eq:final}
\end{equation}
\noindent
Apart from an irrelevant shift of the energy zero and a renormalization
of the frequency into $\omega = \omega_0 (1-\frac{s+t}{2})$
this Hamiltonian contains a quadratic anharmonicity
unless $s=-t$, i.e., unless $p=q^{-1}$.
It is true that a quadratic anharmonicity will emerge even for the
P-type oscillator
($p= q^{-1}$) as a residue of the fourth order term, but
it will be associated with a fourth order anharmonicity which may
well be inconsistent with the experimental spectrum of some
system of interest, such as a diatomic molecule.
 
We finally point out that a coordinate and a conjugate
momentum can be defined for the deformed oscillator by means of
the relations
$\hat{x}= ( a^{\dagger} + a)/\sqrt{2}$ and
$\hat{p}= i( a^{\dagger} - a)/\sqrt{2}$. Application of
Eq.~(\ref{eq:e13}) with the choice $Q=1$ results in 
(for $\hbar \omega_0 = 1$) 
\begin{equation}
[\hat{x}, \hat{p}] = i[a, a^{\dagger}]=
i \lbrack 1+\Big{(}s+t-\frac{(s+t)^2}{2} \Big{)} \hat{n} +
\frac{s^2+st+t^2}{2} \hat{n}(\hat{n}+1) + \cdots \rbrack \; ,
\end{equation}
\noindent
from which follows the deformed uncertainty relation.

\bigskip
\noindent
{\bf Acknowledgements}
 
\noindent
One of the authors (JK) would like to thank the
R\'{e}gion Rh\^{o}ne-Alpes for a visiting fellowship
and the Institut
de Physique Nucl\'{e}aire de Lyon for its kind hospitality.

\end{document}